# Asymmetric 4.77 Three-Way Unequal Filtering Power Divider/Combiner for Communication Systems Application

Augustine O. Nwajana *, Mosammat Rokaiya Akter, and Muhammad Asfar Saeed

University of Greenwich, School of Engineering, Medway Campus, Kent, ME4 4TB, UK

* Correspondence : a.o.nwajana@greenwich.ac.uk

## *Abstract*

*This study presents a novel three-way unequal filtering power divider/combiner, addressing challenges in unequal power distribution while incorporating filtering functions in communication systems. Wilkinson power divider (WPD) is the traditional power division approach using quarter-wavelength transmission lines [1]. This type of power divider is popularly used in communication systems due to its good electrical isolation and simple structure. The problem with WPD is that its operation requires the use of an externally connected bandpass filter (BPF) to achieve filtering functionality. This leads to increased footprint and increased loss coefficients in a system. In contrast to the traditional design approach involving a BPF, a matching transmission line, and a Wilkinson power divider as separate components, the proposed integrated filtering power divider (FPD) consolidates all three components into a single device, leading to lower footprint and lower loss coefficient in a system. Circuit modelling and electromagnetic (EM) simulations were conducted to ensure alignment between theoretical and practical results. The design demonstrates effective unequal power division at the three output ports while maintaining very good filtering performance. Results show a return loss better than 15 dB and a minimum insertion loss of 1.2 dB. The overall size of the device is 32.2 x 50.0 mm. This paper contributes to advancements in power divider design by addressing unequal power division challenges and integrating filtering functions. The findings offer a foundation for future developments in advanced power divider/combiner systems, with insights into potential challenges and areas for further improvements.*

***Keywords:*** *3-way; bandpass filter; FPD, microstrip; power combiner; power divider; SOLR, square open-loop resonator*

## 1. Introduction

In modern telecommunications and signal processing systems, the efficient management of power distribution and filtering is crucial. Power dividers are critical components in wireless communication systems, enabling the distribution of an input signal to multiple paths without significant losses. This efficient power distribution is essential for systems where a single source needs to drive multiple components such as antennas, amplifiers, or other subsystems. Power dividers like the Wilkinson power divider (WPD) [1]–[3] provide excellent isolation between output ports, preventing crosstalk and interference. While the WPD excels in providing equal power distribution and high isolation, it has limitations when unequal power distribution or integrated filtering is required. Three-way (3-way) unequal filtering power divider (FPD) overcomes these limitations by offering significant advantages. It distributes power unequally among its three output ports while simultaneously providing signal filtering capabilities within a specific frequency range. This feature is critical for applications that require specific power ratios for different components, optimizing performance and fulfilling unique system requirements. FPDs achieve power division and filtering capabilities in a single subsystem as reported in [4]–[16]. This integration eliminates the need for separate filters, reducing the physical footprint, cost, and insertion losses associated with additional components and connectors. The integrated filters ensure that only the desired frequency bands are passed, enhancing signal integrity and system performance. The WPD can be adapted to provide unequal power division by modifying the impedance values of the transmission lines and the isolation resistor. To achieve unequal power division, the characteristic impedances of the transmission lines connected to the output ports of the WPD need to be adjusted according to the desired power ratios as shown in Figure 1, resulting in impedances Z1, Z2, and Z3. However, adjusting these impedance values away from the standard Z0 (typically 50 Ohms) can lead to impedance mismatches. These mismatches can increase reflection losses at the ports, thereby reducing the overall efficiency of the power divider. Additionally, changes in impedance and the design of the isolation resistor can affect the system's noise figure. In radiofrequency (RF) and microwave circuits, impedance mismatches can introduce additional noise, which can further degrade the signal-to-noise ratio (SNR) of the system.

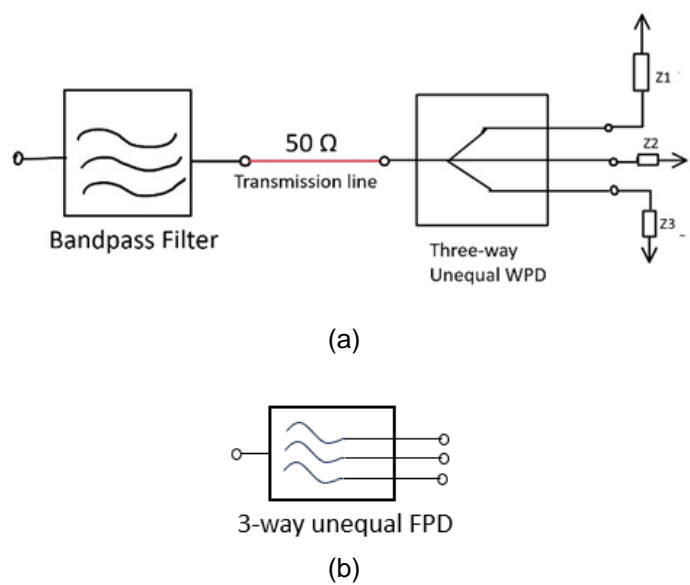

(a)

(b)

**Figure 1**. 3-way FPD design: (a) Traditional cascaded approach; (b) proposed design [2].

## 2. Related Work

The field of microwave engineering and telecommunications has seen significant advancements with the development of power dividers/combiners, which are critical components in various high-frequency applications. These devices play a crucial role in signal processing, wireless communication, and radar systems by distributing power among multiple paths. Traditional power dividers typically focus on equal power distribution; however, there is a growing need for designs that accommodate unequal power division to meet the demands of advanced communication systems.

Researchers have recently reported 3-way FPDs realized using various design techniques. Zhang et al. [4] proposed a wideband 3-way FPD developed on a single multimodal patch microstrip resonator. Though the proposed design is complex due to the handling of the multimode triangular resonator, but it however achieved compactness due to the use of just a single resonator in the design. Chen et al. [6] reported a wideband 3-way FPD with unequal power division ration. The design utilized waveguide technology for design implementation leading to a bulky subsystem, though with good power handling capability. Zhu et al. [7] presented two waveguide FPDs with flexible power division ratio and enhanced selectivity. One of the reported FPDs achieved equal power division ratio of 1:1:1 while the other achieved an unequal power division ratio of 1:2:1. The presented equal and unequal FPDs operated at centre frequencies of 3.7 and 4.5 GHz, respectively. Liu et al. [9] employed unequal-width three coupled lines in the development of a wideband 3-way FPD. The design was composed of short-circuited stubs loaded unequal power divider, isolation resistors, and unequal-width three-coupled lines, and operated at a centre frequency of 1.5 GHz. Qian and Pang [11] implemented a 3-way FPD utilizing the twenty-first century substrate-integrated waveguide technology. The design achieved both in-phase and anti-phase outputs and operated at the centre frequency of 6.41 GHz. In another research article, Nwajana and Paul [12] reported a 3-way FPD implemented using compact microstrip folded-arms square open-loop resonator. The design achieved a minimum return loss of 15.5 dB, an insertion loss of 4.77+0.34, and an isolation of greater than 10.1 dB. Guo et al. [13] proposed a compact 3-way FPD based on three-line coupled structure with a good isolation greater than 20.4 dB. The design achieved a power division ratio of 1:1:2 and operated at a centre frequency of 3.5 GHz. Zhang et al. [14] investigated a general method for developing multi-way FPDs for industrial applications. The proposed new design technique achieved high isolation, good port matching, and compact size. Nwajana and Paul [15] reported a 3-way equal power division ratio FPD. The design is centred at 2.6 GHz frequency with a minimum return loss of 15.5 dB, and port isolation of 0.34 dB. Zhu et al. [16] developed 3-way FPD with harmonic suppression. The design is based on a triangular patch resonator and achieved a minimum insertion loss of 4.77+1.05, an isolation of better than 19 dB, and an impedance bandwidth of 4.7%.

This paper proposes a compact 3-way FPD that delivers unequal amount of power based on the power division ratio of 1:2:4 to the three output ports, while at the same time attenuating unwanted frequency components. Hence, the proposed integrated FPD acts as a three-in-one device since it consolidates three components (a bandpass filter, a matching transmission line, and a Wilkinson power divider) into a single device as shown in Figure 1. The proposed device removes the complexity of having three separate devices in a communication system by replacing all three with a single device, leading to reduced footprint in the system.

## 3. Schematic Circuit

The schematic circuit configuration for the proposed integrated FPD is achieved by first designing three identical bandpass filters using the standard normalised Chebyshev lowpass prototype filter element-values of $g_0 = g_4 = 1.0$, $g_1 = g_3 = 0.8516$, and $g_2 = 1.1032$. The design centre frequency is chosen as 2.6 GHz, with a fractional bandwidth (FBW) of 0.03, and the input/output characteristic impedance ($Z_0$) of 50 Ohms. The bandwidth design constraint or requirement for the BPF employed in forming the proposed FPD is fixed at three percent (3%) of the specified centre frequency. Hence, the BPF design method employed in this research work considers network and component values that meet the required bandwidth constraint. Therefore, the proposed FPD bandwidth is only controlled during the component BPF design phase. The identical BPF circuits were designed based on the technique reported in [17]–[19] and then coupled together to achieve the integrated FPD circuit arrangement shown in Figure 2(a). The simulation results of the schematic circuit arrangement for the proposed 3-way unequal FPD is shown in Figure 2(b).

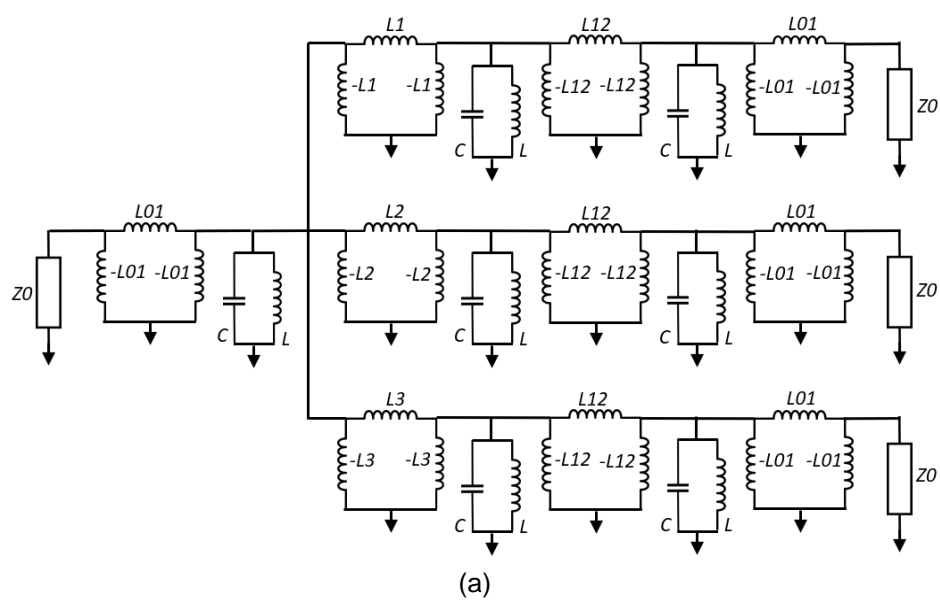

(a)

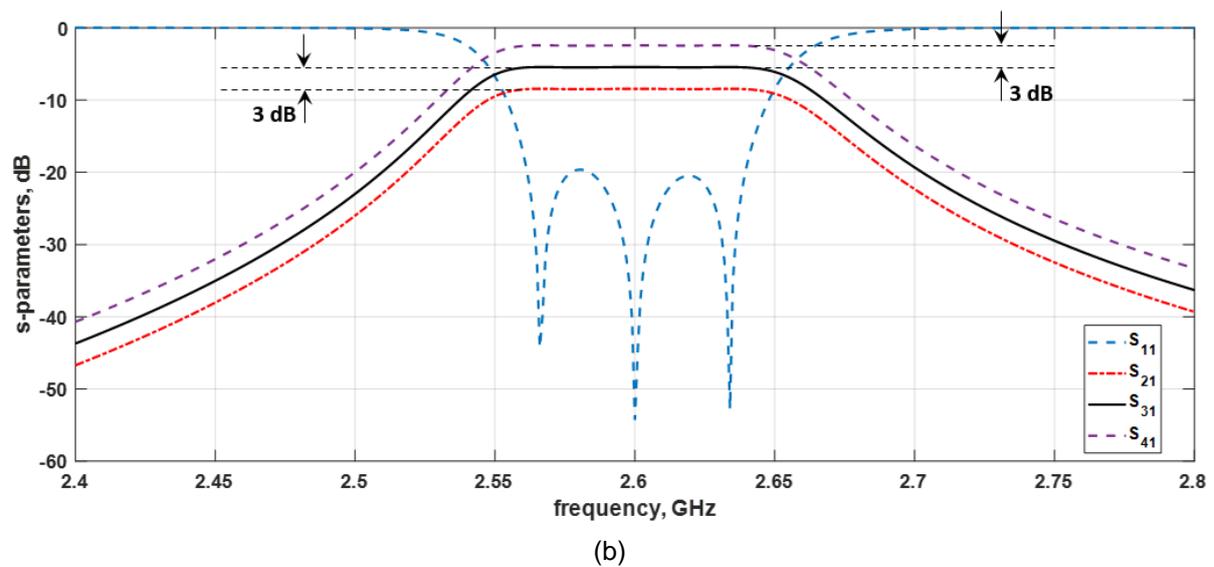

**Figure 2.** Proposed 3-way unequal FPD: (a) Schematic circuit (Z0 = 50 Ω, L = 0.1078 nH, C = 34.7529 pF, L01 = 1.2243 pF, L12 = 1.0756 pF, L1 = 9.2167 nH, L2 = 6.5172 nH, L3 = 4.6084 nH; (b) Simulation results.

## 4. Microstrip Layout

The unequal FPD schematic circuit was implemented using microstrip technology and the square open-loop resonator (SOLR) was chosen for the layout implementation. The width and guided wavelength for designing the SOLR are determined from [19]. Advanced Design System (ADS) EM Simulation Software was used to create and simulate the unequal FPD layout model, with the resonators coupling arrangement shown in Figure 3. The layout was created on Rogers RT/Duroid 6010LM substrate with a dielectric constant of 10.7, a thickness of 1.27 mm, and a loss tangent of 0.0023. The SOLR is designed to resonate at the circuit model specified centre frequency of 2.6 GHz.

As shown in the coupling scheme of Figure 3, it is necessary to determine four mutual coupling coefficient values M23, M1, M2, and M3 using Equations (1), (2), (3), and (4), respectively. M23 is the mutual coupling coefficient between each resonator pair in the three identical BPFs. M1, M2, and M3 are the mutual coupling coefficients between the common resonator (CR) and the first, second, and third BPFs, respectively. It is important to note that CR contributes three poles to the FPD response as it replaces one resonator from each of the three identical BPFs. This means that CR contributes to the reduced physical footprint of the proposed FPD device. The external quality factor (Qe) is the coupling between the input port (that is, port 1) and CR. It is also the coupling between the last resonator of each BPF and its corresponding output port (that is, ports 2, 3, or 4) indicated in Figure 3. The coupling value for Qe is determined using Equation (5) which is derived from [17–19].

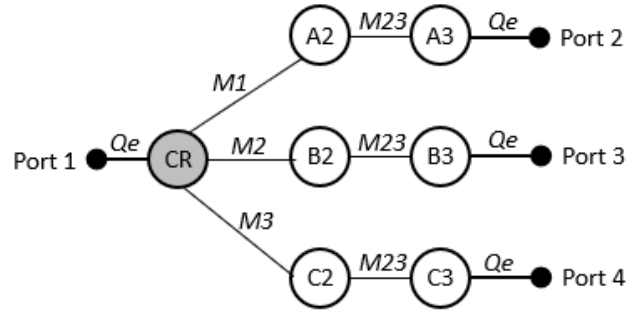

**Figure 3.** Coupling arrangement for the proposed 3-way unequal FPD.

$$M_{23} = \frac{FBW}{\sqrt{g_1 g_2}} = 0.031 \tag{1}$$

$$M_1 = \frac{M_{23}}{\sqrt{7}} = 0.012 \tag{2}$$

$$M_2 = \frac{\sqrt{2} M_{23}}{\sqrt{7}} = 0.017 \tag{3}$$

$$M_3 = \frac{2 M_{23}}{\sqrt{7}} = 0.023 \tag{4}$$

$$Q_e = \frac{g_0 g_1}{FBW} = 28.387 \tag{5}$$

The physical dimensions of the unequal FPD are established using full-wave Keysight ADS electromagnetic simulator which simulated the mutual coupling coefficients and the external quality factor against physical structures as reported in [12]. The physical dimensions of the 3-way unequal FPD microstrip layout are shown in Figure 4. The overall physical size (excluding the input/output connectors) of the fabricated unequal FPD device is 32.2 by 50.0 mm.

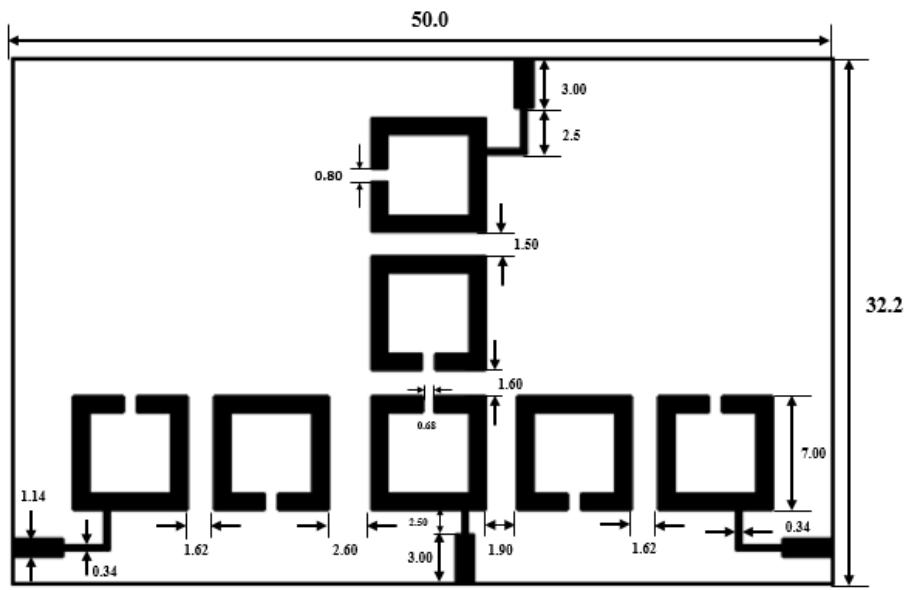

**Figure 4.** Microstrip layout of the 3-way unequal FPD.

## 5. Results and Discussion

This section captures and discusses both the schematic circuit and layout simulation results for the proposed unequal FPD. The results are jointly presented in Figure 5 for ease of analysis and comparison. The proposed unequal FPD performance captured in Figure 5 shows good agreement between the schematic circuit and layout responses. The solid line plots in the Figure 5 represent the schematic circuit results while the dashed line plots denote the layout results. Looking at the differential mode responses of Figure 5(a), a layout minimum return loss of 15.0 dB is achieved compared to the 20dB return loss of the schematic circuit results. The achieved layout minimum insertion loss is recorded as 4.77+1.2 dB, where the 4.77 dB part is the standard theoretical inherent characteristics return loss of any 3-way unequal FPD device as reported in [8]. The single-ended output ports responses ($S_{22}$, $S_{33}$, $S_{44}$) shown in Figure 5(b) also confirms the good agreement between the schematic circuit and layout results of the proposed unequal FPD. The schematic circuit and the layout isolation responses ($S_{23}$, $S_{24}$, and $S_{34}$) of the proposed 3-way unequal FPD are shown in Figure 5(c). The proposed design achieved an in-band ports isolation of better than 10.6 dB across all bands.

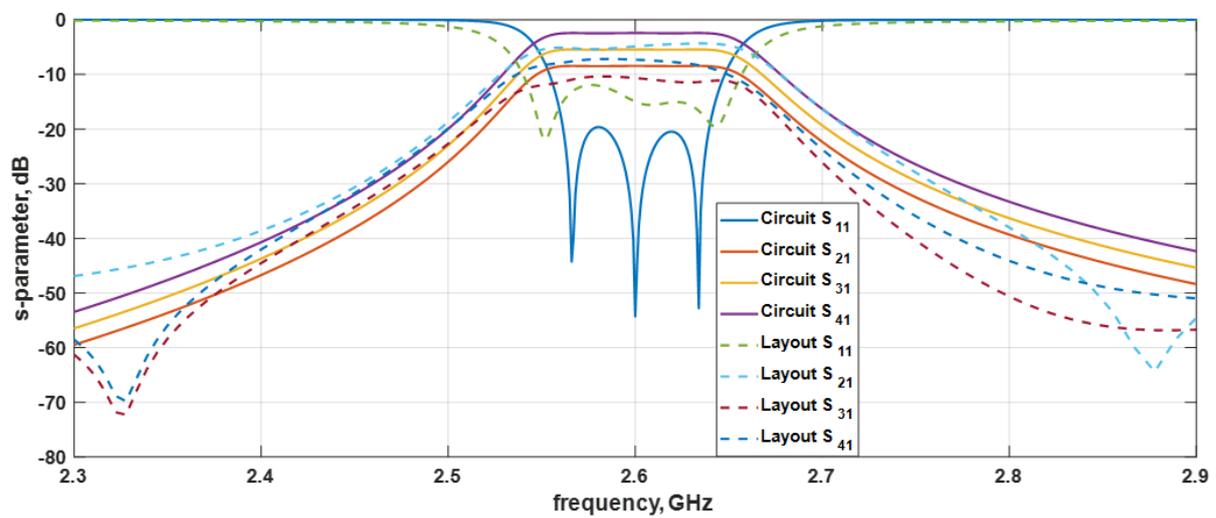

(a)

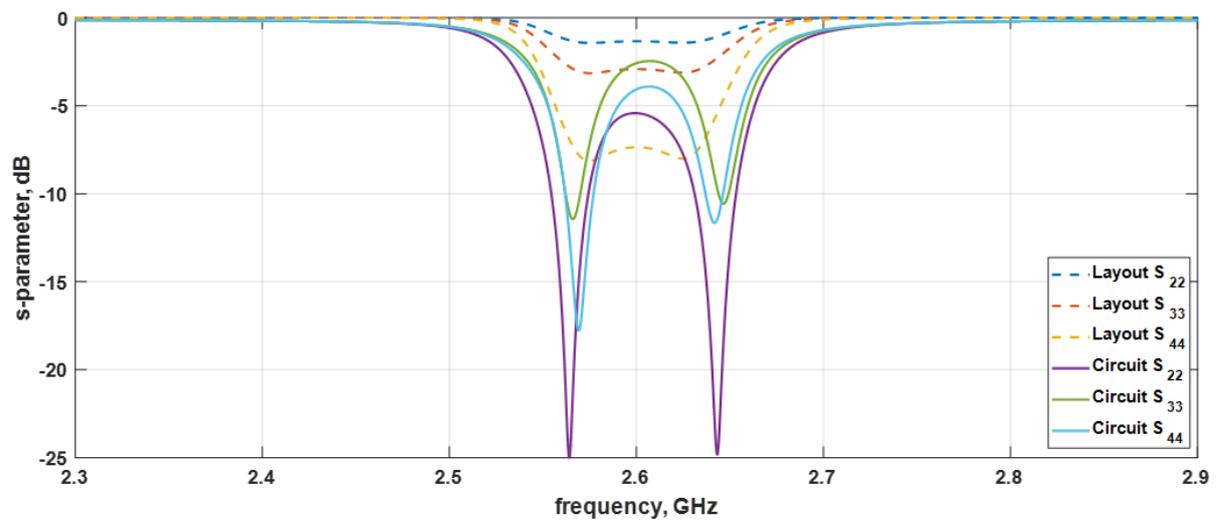

(b)

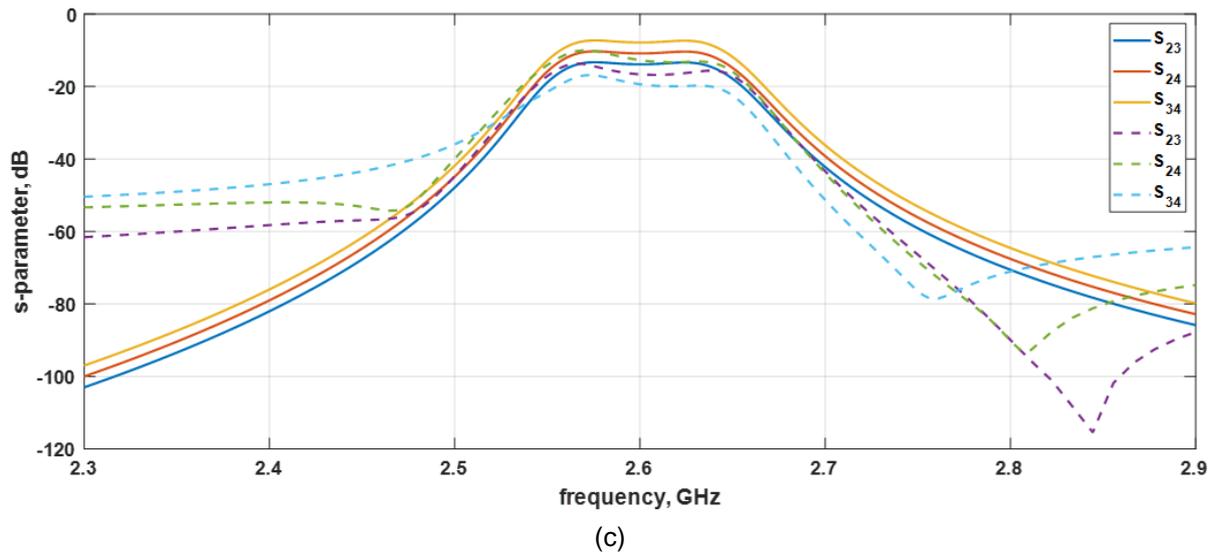

(c)

**Figure 5.** Schematic circuit and layout results comparison: (a) Differential mode results; (b) Single-ended output port results; (c) In-band ports isolation responses.

## 5. Conclusions

A 3-way FPD that receives an input power signal and delivers unequal share of power signals to its three output ports has been developed and realized using microstrip technology. The design implementation is based on the compact SOLR for device miniaturization. The proposed FPD operates at a 2.6 GHz centre frequency, with a 3% fractional bandwidth. The design realization is carried out on Rogers RT/Duroid 6010LM substrate with a dielectric constant of 10.7, a thickness of 1.27 mm and a loss tangent of 0.0023. The schematic circuit and layout results of the proposed FPD indicate good agreement with a return loss of better than 15.0 dB, an insertion loss of better than 4.77+1.2 dB, and an in-band port isolation of better than 10.6 dB. The device achieved a compact size of 32.2 x 50.0. The compact design is well-suited for modern multifunctional communication systems due to its low insertion and return losses, sharp selectivity, and integrated filtering capability.